**Optoelectronic Physical Unclonable Functions and Reservoir-Inspired Computation with Low Symmetry Integrated Photonics**

Farhan Bin Tarik[1,2], Yingjie Lao[3], Mustafa Hammood[4], Jonathan Barnes[4], Madeline Mahanloo[4], Lukas Chrostowski[4], Taufiquar Khan[5], and Judson D. Ryckman[1,*]

[1]*Holcombe Department of Electrical and Computer Engineering, Clemson University, Clemson, SC 29634, USA*
[2]*Department of Electrical, Computer and Cybersecurity Engineering, Florida Polytechnic University, Lakeland, FL 33804, USA (current)*
[3]*Department of Electrical and Computer Engineering, Tufts University, Medford, MA 02155, USA*
[4]*Department of Electrical and Computer Engineering, The University of British Columbia, Vancouver, British Columbia V6T 1Z4, Canada*
[5]*Department of Mathematics and Statistics, University of North Carolina at Charlotte, Charlotte, NC 28223, USA*

*jryckma@clemson.edu

## Abstract

Emerging applications of photonics in computing, sensing, and security increasingly demand complex input-output behaviors, including highly nonlinear transformations of optical signals. Traditional photonic systems rely on highly structured components with symmetric geometries and low-entropy modal responses to achieve predictable and analytically describable behavior. To achieve expressive functionality, this paradigm often requires large networks of fabrication-sensitive interferometers or resonators and substantial hardware error correction to restore deterministic operation. Here we demonstrate an alternative paradigm rooted in low-symmetry, disordered integrated photonic circuits, which provide intrinsically enhanced modal diversity and spectral complexity, enabling highly nonlinear transformations of input signals into information-rich outputs. Our devices, physically unclonable moiré quasicrystal interferometers integrated on a silicon photonics platform, exhibit aperiodic and reconfigurable spectral responses and are characterized by analyticity breaking and erasable mutual information. Using dynamic thermo-optic control to drive their complex spectral dynamics, we demonstrate that these devices function as reconfigurable physical unclonable functions (rPUFs). We also highlight their ability to perform high-dimensional input–output transformations, emulating reservoir-inspired information processing in a compact photonic platform. This work bridges the gap between engineered and natural complexity in photonic systems, revealing new opportunities for scalable, energy-efficient, and information-dense optoelectronics with applications in secure communications, hardware security, advanced sensing, and optical information processing. Our results establish low-symmetry integrated photonics as a powerful resource for complex signal manipulation in photonic systems.

## Introduction

In photonics, structural symmetries play a pivotal role in molding the flow of light and dictating its interaction with matter. Photonic systems with high levels of symmetry—whether translational, rotational, periodic, or mirror symmetry—tend to exhibit well-defined behaviors that are often analytically describable in a compact form. This reflects the limited mode diversity and manageable number of interacting degrees of freedom present in highly symmetric optics. However, this simplicity comes with a trade-off: devices designed with high symmetry tend to lack the complexity and expressivity needed for advanced functionalities. One dominant vector for addressing this problem, is to cascade symmetric components together into larger and more complex circuits and systems[1–10]. This effectively increases the mode diversity, functionality, and expressivity of optical circuits enabling more advanced and complex behaviors. While these properties are favorable for applications including optical computing, communications, sensing, and hardware security – they also present challenges in design, footprint, efficiency, and the need for precise control and error correction[11–13].

In contrast, photonic systems with broken or low structural symmetry open new opportunities for achieving complex behaviors by supporting intrinsically enhanced mode-diversity. Such structures can support complex multi-mode resonance patterns, diverse interferometric or spectral responses[14–16], and transitions between Anderson localization and delocalization which collectively create complex input-output behaviors[17]. Unlike symmetric systems with engineered complexity, the behavior of low symmetry optics with natural complexity often cannot be compressed into a compact analytical description, thus necessitating advanced numerical methods to computationally emulate or approximate system behaviors. Although challenging to model, the enriched complexity of such systems may provide an inherent resource for emerging applications which demand complex and nonlinear transformation of input signals into output signals in a compact and efficient platform.

The distinctions between high vs. low symmetry photonics and engineered vs. natural complexity are also mirrored in other scientific domains. A primary example of this is the distinction between conventional deep neural networks vs. random neural networks which rely on engineered complexity (training internal weights) vs. natural complexity (random internal weights) respectively. Recently, it has been shown that entirely linear optics can perform complex nonlinear information processing[18–21]. In effect, photonic devices/systems can be computationally interpreted as the physical manifestation of a neural network. While certain applications may demand high dimensional signal transformation to facilitate efficient data processing or pre-processing, others may demand an unclonable or computationally challenging task. These requirements are not always mutually exclusive. One recent demonstration has shown how random multiple-scattering in a dynamically modulated integrating sphere can perform both functions simultaneously[19,22], effectively rendering capabilities akin to a steady-state reservoir computer. Translating these characteristics to an integrated photonics platform is crucial for realizing compact, scalable, energy-efficient, and high-performance solutions in fields such as hardware security, optical computing, and advanced sensing – but has so far remained largely unexplored.

Here, we report on the dynamic control of low symmetry photonic integrated circuits. Our devices are comprised of physically unclonable moiré quasicrystal interferometers (QCIs)[15,23] realized in a silicon photonic platform with integrated thermo-optic heaters controlled by an externally applied bias[24]. These devices are shown to exhibit a range of distinctive features typically inaccessible to high symmetry photonics, including aperiodic and reconfigurable spectral responses with erasable mutual information, analyticity breaking, and spectrally multiplexed neuron behaviors that are both highly expressive and unclonable. We also demonstrate how the high-dimensional input-output signal transformation can be harnessed to perform reservoir-inspired processing. This work points toward new, scalable applications of long-studied localization and delocalization wave phenomena arising in disordered media and quasicrystals[25–27]. More generally these results establish low-symmetry integrated photonics as a promising platform for optoelectronic information processing, with prospective applications in hardware security, sensing, communications, and optical computing.

## The optoelectronic quasicrystal interferometer

The QCI (Fig. 1) is a specialized multiple-scattering interferometer, conceptually similar to a Michelson–Gires–Tournois interferometer but with the front mirrors replaced by 500 µm-long moiré quasicrystals[15]. These quasicrystals are realized using width-modulated single-mode silicon nanowire waveguides [28,29], formed by superimposing gratings with 316 nm and 317 nm periods as described in prior work[15]. This design yields a ~100 µm moiré beat length and a spectral region near 1550 nm that exhibits localization phenomena (Fig. 2). This region, akin to a mobility gap, supports high-Q localized states ($Q > 10^4$) as well as Bragg-like low-transmission behavior. Adjacent spectral bands support delocalized states with low group velocity and complex longitudinal mode structure. In this slow-light, band-edge-like regime, even minute geometry fluctuations accumulate enhanced phase delays, making the optical response exceptionally sensitive to fabrication disorder. At the same time, the extended longitudinal mode structures sample a large spatial region of the device, causing the spectrum to depend intricately on the specific distribution of those imperfections. Rather than simply accumulating phase error as in a conventional interferometer, these imperfections (Fig. 1b, c) form a distributed multiple-scattering potential that supports a rich set of localized and delocalized modes that contribute to the QCI's spectrally diverse and physically unclonable resonant response[15,23].

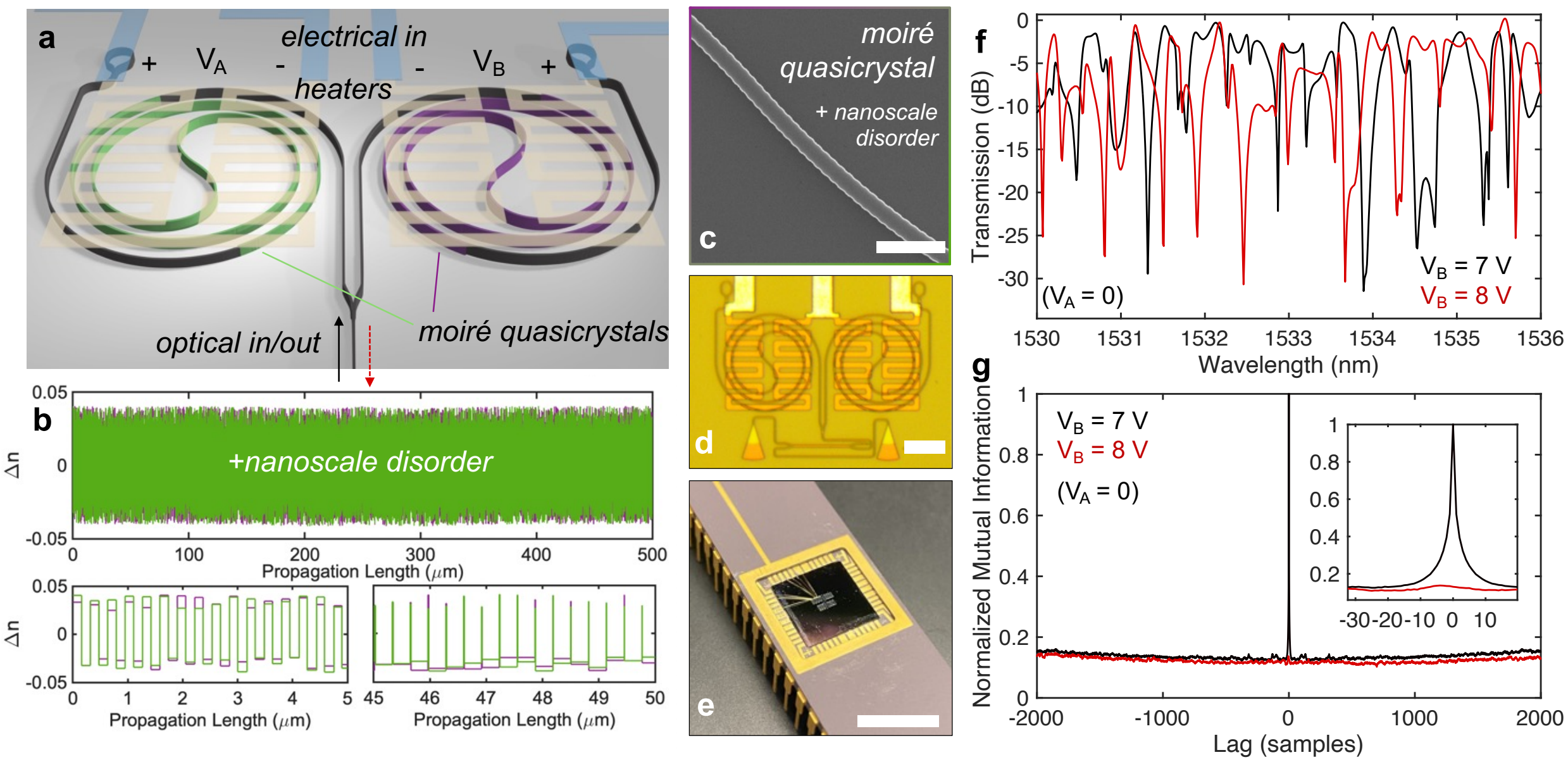

**Figure 1. The optoelectronic quasicrystal interferometer (QCI).** (a) Illustration of the QCI with resistive heaters above each arm. Each arm contains a 500 µm long waveguide quasicrystal (green/purple regions), a 150 µm long normal waveguide, and a loop mirror. (b) Illustration of 1D quasicrystal potentials capturing fabrication induced disorder. (c) SEM image of a waveguide quasicrystal. (d) Microscope image of a prototype QCI. (e) Photograph of a prototype QCI chip after wirebonding. (f) Transmission spectra and corresponding normalized AMI and CMI for $V_B = 7$ and 8 V ($V_A = 0$).

Dynamic control of the QCI is achieved thermo-optically using Ti/W alloy resistive heaters integrated above each quasicrystal (Fig. 1d). Unlike conventional interferometers, which exhibit periodic spectra that shift predictably with applied voltage, the QCI produces an aperiodic spectrum that evolves deterministically but in a more complex and less predictable fashion. Figure 1f illustrates this behavior near the mobility edge, where modulating one arm from 7 V to 8 V results in a nontrivial spectral change.

We performed auto mutual information (AMI) and cross mutual information (CMI) analysis on the device spectra (Fig. 1e) [30]. Unlike correlation-based methods (e.g. auto-correlation, cross-correlation) that detect only linear dependencies, mutual information captures both linear and nonlinear relationships, making it particularly well suited for analyzing complex and potentially non-analytic spectral patterns. The presence of a single peak in the AMI confirms the aperiodic nature of the QCI spectrum, in contrast to traditional interferometers or ring resonators, which exhibit multiple peaks due to their well-defined free spectral range (Figs. S1 & S2)[31]. The CMI analysis reveals that the 7 V and 8 V spectra are not related by a simple spectral shift, but instead represent distinct spectral fingerprints. This result supports the use of the QCI as a reconfigurable physical unclonable function (rPUF)[32,33], where a digital key can be derived from the device's optically encoded spectral response.

To shed greater insight into the operation of our QCI device, we performed numerical simulations of light transmission through an individual quasicrystal. Figure 2A shows the spatially resolved electric field intensity $|E|^2$ across wavelength, revealing transitions between delocalized and localized optical states. Localized modes appear in well-defined spatial regions determined by the moiré superlattice, with a characteristic beat length of ~100 μm (indicated by dashed lines), reflecting the quasi-deterministic nature of these states. In contrast, delocalized modes near the band edges exhibit complex, extended field distributions across the device, as visualized in Fig. 2B. The delocalized states near the spectral transition region are expected to contribute most strongly to device entropy, as they are exquisitely sensitive to distributed fabrication imperfections. These results provide theoretical support for the QCI's rich spectral behavior and help explain its capacity for complex and reconfigurable optical responses. When integrated into interferometric circuits with loop mirrors and dual optical paths, these building blocks give rise to complex, aperiodic, reconfigurable, and physically unclonable optical responses.

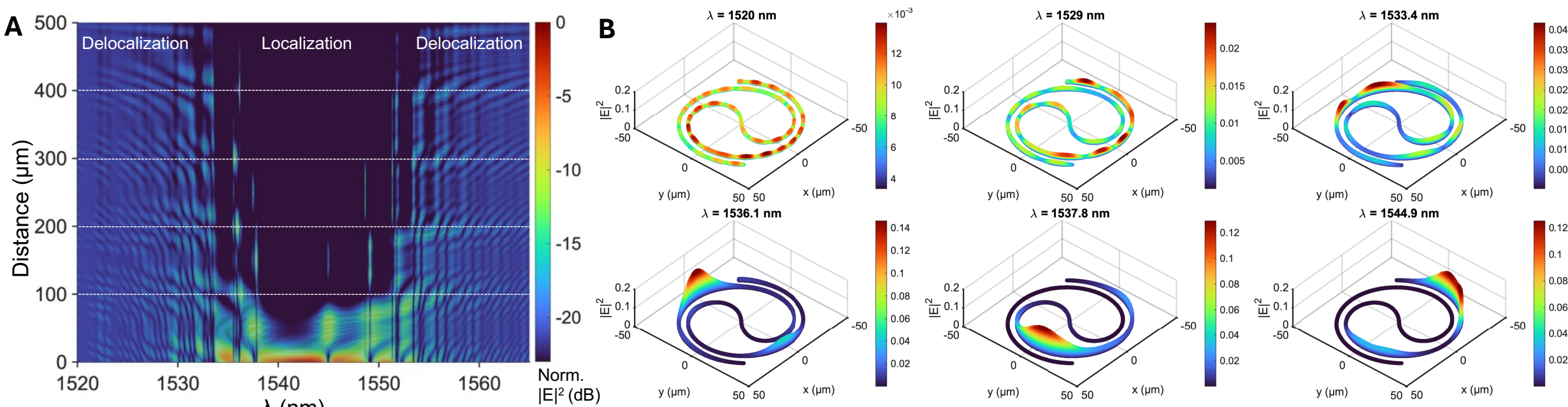

**Figure 2. Simulated light transport in an individual moiré quasicrystal.** (A) Spatially resolved electric field intensity $|E|^2$ as a function of wavelength, showing transitions between delocalized and localized optical states. Localized modes emerge at quasi-deterministic positions defined by the moiré superlattice, with a characteristic beat length of ~100 μm (indicated by dashed lines). (B) Electric field intensity profiles at selected wavelengths spanning the delocalized-to-localized transition. Delocalized modes exhibit complex, extended distributions, while localized modes are confined to distinct regions.

## Dynamic reconfiguration of spectral fingerprints

To study the non-trivial interferometric response of the QCI we recorded coarse (0 to 15 V, 0.5 V increment) and fine (7 to 8 V, 0.05 V increment) voltage scans across three identically designed QCI rPUFs. Coarse scans for one device are shown in Fig. 3a,b, illustrating the dynamic spectral evolution near the edges of the mobility gap. The spectra reveal a dense set of resonant features that red-shift in proportion to the applied heater power ($\propto V^2/R$), while exhibiting a complex and voltage-dependent interplay of phase relations between multiple modes. Heating a quasicrystal effectively redshifts its associated mobility edges, while the other quasicrystal exhibits a weaker but non-zero spectral shift due to thermal crosstalk. At a fixed operating wavelength, this enables active modulation of the interference landscape: the QCI can selectively suppress or enable localized states and dynamically reshape the delocalized mode profiles, effectively reprogramming the optical transport characteristics of the structure.

Figure 3c compares the dynamic spectral response of three QCI rPUFs as one arm is modulated from 7 V to 8 V. Additional data is reported in Fig. S3. Each device exhibits a distinct spectral fingerprint, and more importantly, reconfigures into a new fingerprint under electrical tuning. Intra-device CMI analysis (Fig. 3e) quantifies the reconfigurability of a single rPUF, while inter-device CMI analysis (Fig. 3f) confirms the spectral uniqueness between different rPUFs.

Each pixel in the computed CMI($V_i$,$V_j$) measures how predictable the spectrum at voltage $V_i$ is given the spectrum at voltage $V_j$. The bright diagonal and dark off-diagonal regions of Fig. 3e indicate the spectra for $V_i \neq V_j$ are unique and not related by a simple spectral shift. In contrast, the absence of a bright diagonal in Fig. 3f confirms that two different rPUFs are uncorrelated. Faint streaks in Fig. 3e,f indicate the presence of weakly correlated spectral features that shift with voltage, with residual CMI generally below 0.25. The intra-device results (Fig. 3g and Fig. S4) show that mutual information drops as the heater power is modulated, with ~1.25 mW of power required to reconfigure the spectrum and reduce the normalized mutual information below 0.5, a representative threshold below which spectral features are sufficiently decorrelated to be considered distinct configurations.

This level of modulation implies that a QCI rPUF driven by a single 0-15 V electrical input can be reconfigured into ~$10^2$ distinct optical keys. With dual-input modulation, the accessible key space expands to ~$10^4$ unique keys. Assuming a 2048-bit key length—comparable to our prior work[15]—this corresponds to ~20 Mb of total key material encoded in a single QCI. Given that the active information-bearing region comprises only ~$7\times10^{-6}$ $cm^2$ of silicon waveguides, the resulting information density exceeds 10 $Gb/mm^2$. Although the realized density is reduced in practice due to the footprint of routing and heaters, this result highlights the remarkable information capacity and functional richness of low-symmetry photonic systems. In the future, this information density could be scaled by several orders of magnitude using more compact device architectures, broader spectral bandwidth, polarization diversity, leveraging optical nonlinearities, and/or by increasing the number of input and output channels.

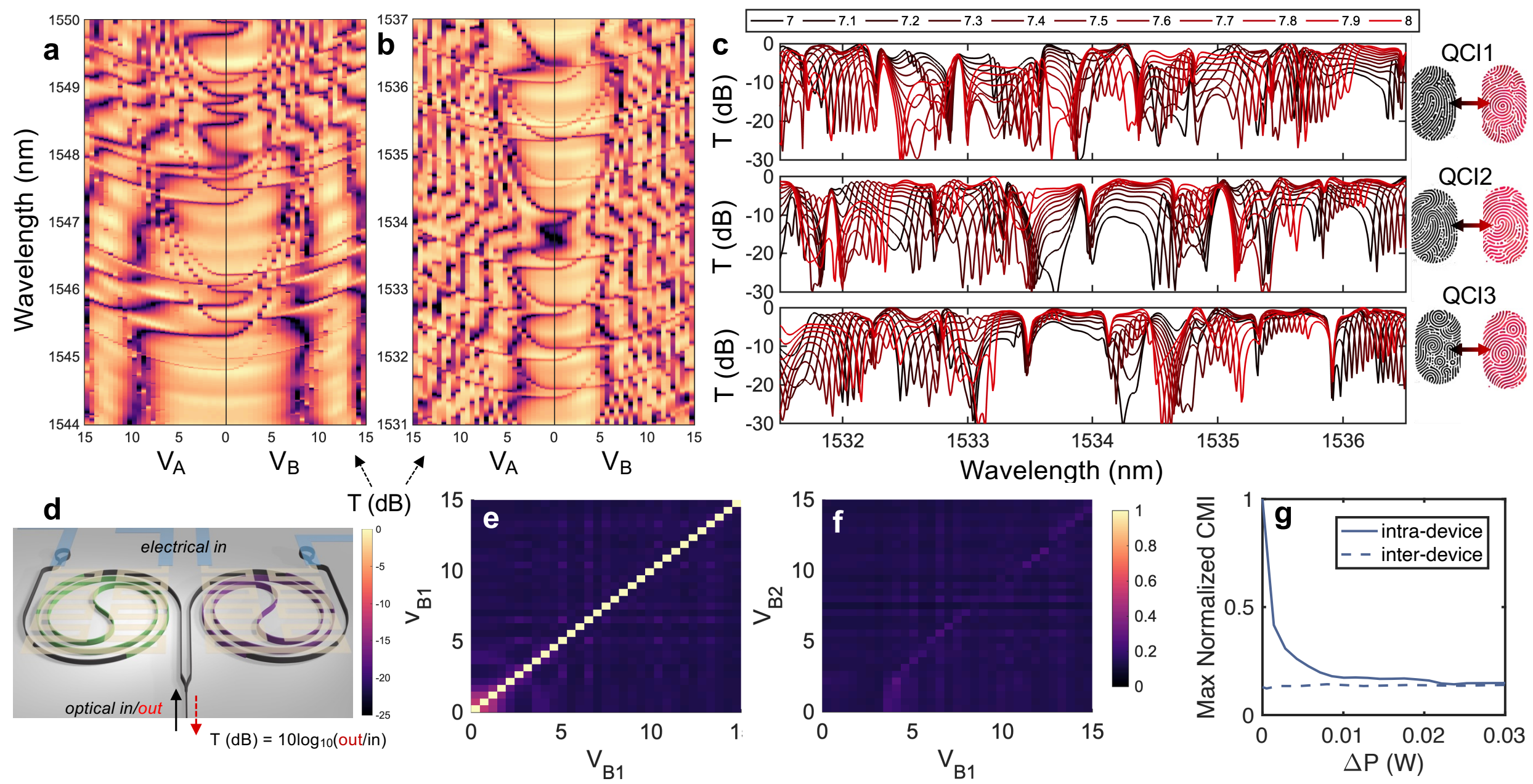

**Figure 3. Spectral characterization, reconfigurability, and unclonability.** (a, b) Transmission spectra (dB scale) for a QCI vs. applied voltage. (c) Transmission spectra for three different QCI rPUFs as $V_B$ is varied from 7 V to 8 V ($V_A = 0$), revealing reconfigurable device fingerprints. (d) Device illustration and definition of transmission. (e) Max intra-device and (f) inter-device normalized mutual information (NMI) for vs. applied voltage. (g) Summary of the intra-device and inter-device max normalized CMI vs. applied heater power.

## State-space dynamics, analyticity, and expressivity

Expressivity requires a system to generate a wide range of distinct, non-redundant responses across its input space, enabling diverse and nonlinear transformations from input to output. To assess the potential expressivity of the QCI more closely, we examined its dynamic spectral behavior under thermo-optic modulation and compared it to that of a conventional ring resonator. We visualized these dynamics using state-space embeddings (Fig. 4e–h), where each trajectory represents the transmission at a fixed wavelength across three successive voltages: T(V), T(V+δV), and T(V+2δV).

The ring resonator (Fig. 4e) traces a simple approximately closed loop trajectory — indicative of a low-dimensional, analytic system. In contrast, the QCI (Fig. 4f–h) traces non-repeating curves with complex internal structure and minimal redundancy. Each incremental voltage shift induces a distinct transformation in the system's spectral response, yielding a complex, non-analytic input–output mapping that resists reduction to a simple underlying model.

While time-evolving systems are often analyzed through state-space embeddings of a variable $x(t)$ over successive time delays, revealing attractors or simple governing rules, the QCI operates in a steady-state regime driven by voltage rather than time. Its state-space behavior lacks cyclicality or closed loop structure and does not show evidence of a stable attractor. Instead, our system exhibits characteristics often associated with transient chaos, the most important of which include non-attracting dynamics, irregularity, unpredictability, and an ordered yet intricate state-space structure [34]. Notably, transient chaos is suggested to play a key role in the expressive power of deep neural networks as they compute high dimensional nonlinear mappings[35,36].

To further illustrate this behavior, we created animations that visualize the 2D state-space trajectory T(V) vs. T(V+δV) as the delay δV is gradually varied (see Supporting Videos 1-4). Remarkably, this evolving 2D projection appears to rotate and morph like a three-dimensional object, suggesting an underlying structure that is smooth and deterministic yet high-dimensional. In effect, we are observing a shadow of the manifold on which the device operates—a complex geometric structure shaped by its internal scattering dynamics and nonlinear voltage response. This highlights the expressive richness of the QCI and reinforces its potential as a physical substrate for encoding, computation, or reconfigurable photonic functions.

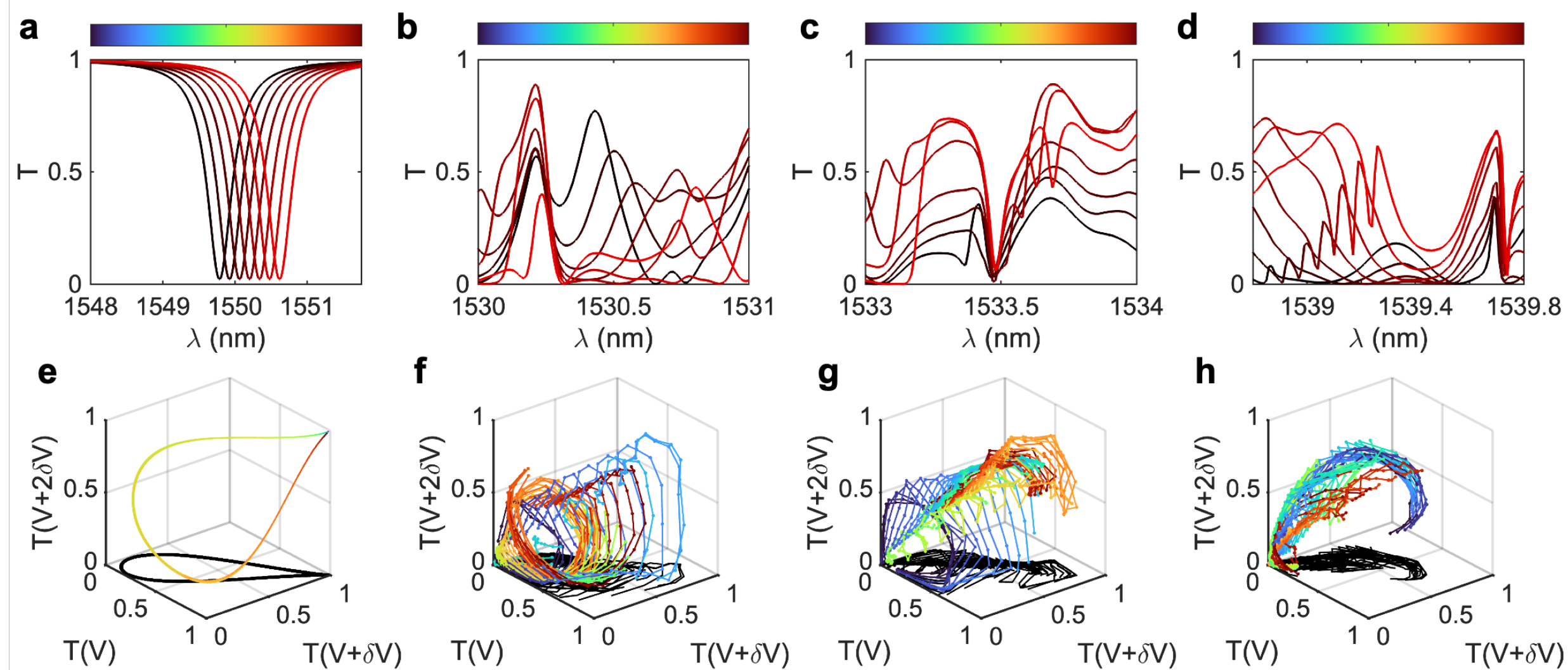

**Figure 4. Complex dynamic spectral behavior visualized in state-space.** (a) Modulated ring resonator spectra (analytic) vs. (b-d) modulated QCI spectra (non-analytic) in selected spectral windows. State-space embedding plots for the corresponding (e) ring resonator and (f-h) QCI spectra, with curves colored according to wavelength, consistent with the color bars in (a–d). A fixed voltage increment δV = 0.1 V is used.

## Steady state reservoir-inspired computing

Lastly, we demonstrate how the complex and expressive behavior of our optoelectronic PUF can be harnessed for nonlinear information processing (Fig. 5). Inspired by reservoir computing, we construct a high-dimensional reservoir state $\mathbf{z} \in \mathbb{R}^N$ from the wavelength-dependent transmission measured at $N$ selected frequencies (wavelengths), such that $\mathbf{z} = [T_1, T_2, \ldots, T_N]^\top$. The PUF nonlinearly maps a set of input and control variables $\boldsymbol{x}$, $\boldsymbol{\omega}$, and $\boldsymbol{\theta}$ to this output state $\mathbf{z}$, where $\boldsymbol{x}$ is the electrical input (stimulus), $\boldsymbol{\omega}$ defines the probe frequencies, and $\boldsymbol{\theta}$ represents the system's hidden internal parameters—comprising fixed structural properties and fabrication-induced disorder that give the PUF its unique fingerprint. In this framework, the PUF effectively implements a random feature map, projecting the inputs into a rich, high-dimensional space where simple readout functions (e.g., linear classifiers or regressors) can be trained to perform complex tasks.

Figure 5b–c shows the evolution of the PUF reservoir states for three QCI devices, where we set $V_A = 0$ and define the single input variable $x = V_B - 7$, which shifts the 7 V to 8 V operating range into the normalized interval $x \in [0,1]$. We focus on the 7 V to 8 V region because it provides representative mid-range spectral dynamics suitable for fine voltage scanning. Figure 5b presents the spectral response of each PUF across the range 1530–1540 nm as $V_B$ varies. From this data, we extract the optoelectronic reservoir features $z_i(x)$ for N = 21 wavelengths as shown in Fig. 5c. Each spectral channel provides a distinct and nonlinear transformation of the input, corresponding to a diverse high-dimensional encoding inherent to each physically unique QCI PUF. Collectively, these transformations form a randomized basis of features with high expressive power, enabling the system to approximate a broad class of functions.

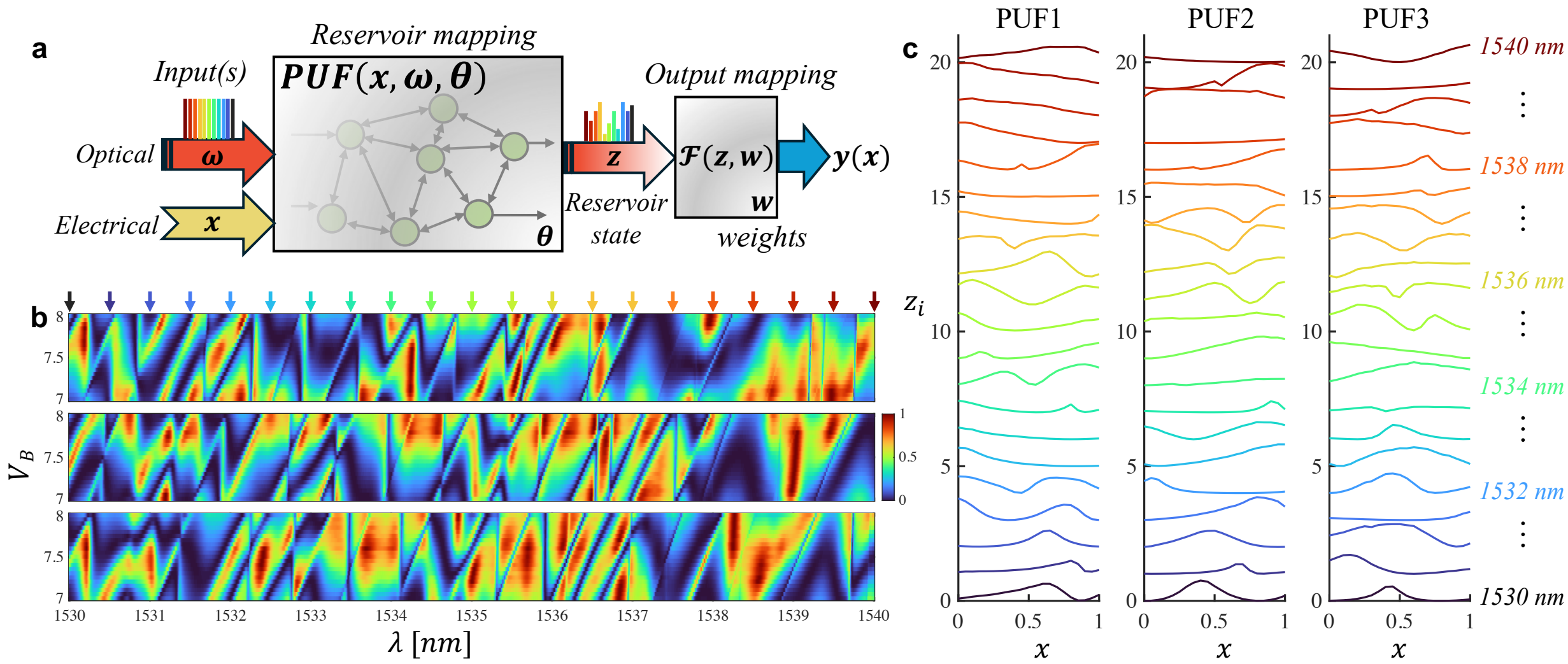

**Figure 5. Physically unclonable nonlinearities.** (a) Schematic of the photonic system mapping electrical input $\boldsymbol{x}$ and optical frequencies $\boldsymbol{\omega}$ to output $\boldsymbol{y}$ via a PUF reservoir $\mathbf{z}$ and learned weights $\boldsymbol{w}$. (b) Spectral response map for three different PUFs. (c) Nonlinear reservoir features $z_i(x)$ for three PUFs ($x = V_B - 7$).

This reservoir state $\mathbf{z}$ can be mapped to an output $\mathbf{y}$ through a function $\boldsymbol{\mathcal{F}}(\mathbf{z},\mathbf{w})$, where $\mathbf{w}$ contains trainable weights used in the output mapping. In conventional reservoir computing, the output function is typically linear—i.e., $\boldsymbol{\mathcal{F}}(\mathbf{z},\mathbf{w}) = \mathbf{w}^{\mathsf{T}}\mathbf{z}$ —enabling efficient training via simple linear or ridge regression. This framework shares conceptual similarities with other classes of randomized neural networks such as extreme learning machines (ELMs)[37,38], in which the internal mapping is fixed and only the output weights are trained. In recent work by Xia et al., the random feature mapping, or reservoir, is referred to as a 'deep nonlinear optical encoder', while the output mapping is implemented as a nonlinear multi-layer perceptron digital decoder via post-processing[19]. In our present system, the output mapping is also performed in post-processing, while the nonlinearity and high dimensional signal transformation is encoded by the QCI PUF. In principle, the output mapping could also be implemented all-optically by weighting the $N$ probe wavelengths and summing the result with a photodetector to enable fully analog inference in a compact photonic platform[39].

Unlike a traditional reservoir computer, which relies on temporal recurrence and satisfies the echo-state property with fading memory, our device operates as a memory-free steady-state reservoir. Internally it exhibits recurrent (multiple) scattering but transforms inputs to outputs in a feed-forward manner. Here we train only the output layer of an equivalent neural network (Fig. 6a), analogous to an ELM[37,38], to demonstrate function approximation. The PUF-derived nonlinear spectral responses provide a randomized set of basis functions $z_i(x)$ which are linearly combined via trained weights to reconstruct a pseudorandom PAM-4 target output waveform (Fig. 6b).

To visualize function approximation quality, we generate an eye diagram by overlaying successive symbol intervals of the output PAM-4 signal on a common $x$-axis (Fig. 6c). An open eye indicates that the four amplitude levels and their transitions (e.g., 0→1, 3→0, etc.) are correctly reproduced. As shown in Fig. 6c, increasing the number of reservoir features $N$ enhances the fidelity of the approximation, though excessive capacity leads to overfitting. Figure 6d plots the approximation accuracy versus $N$ across three different QCI PUF devices, highlighting the expressive power of their distinct, physically encoded nonlinearities.

The results in Fig. 6 confirm the function approximation capability of the proposed neural architecture. Similar results are obtained for other target functions as illustrated in supplementary information Figs. S5 & S6. Although the ELM type neural architecture depicted in Fig. 6 can be illustrated with only one hidden layer, each random nonlinear activation function or reservoir component, $z_i = f_i(x)$, effectively computes a deeper multi-layer perceptron with traditional activation functions. Moreover, each of these functions is reconfigurable by activating the other electrical input, equating to an even more complex reservoir mapping $z_i = f_i(x_1, x_2)$.

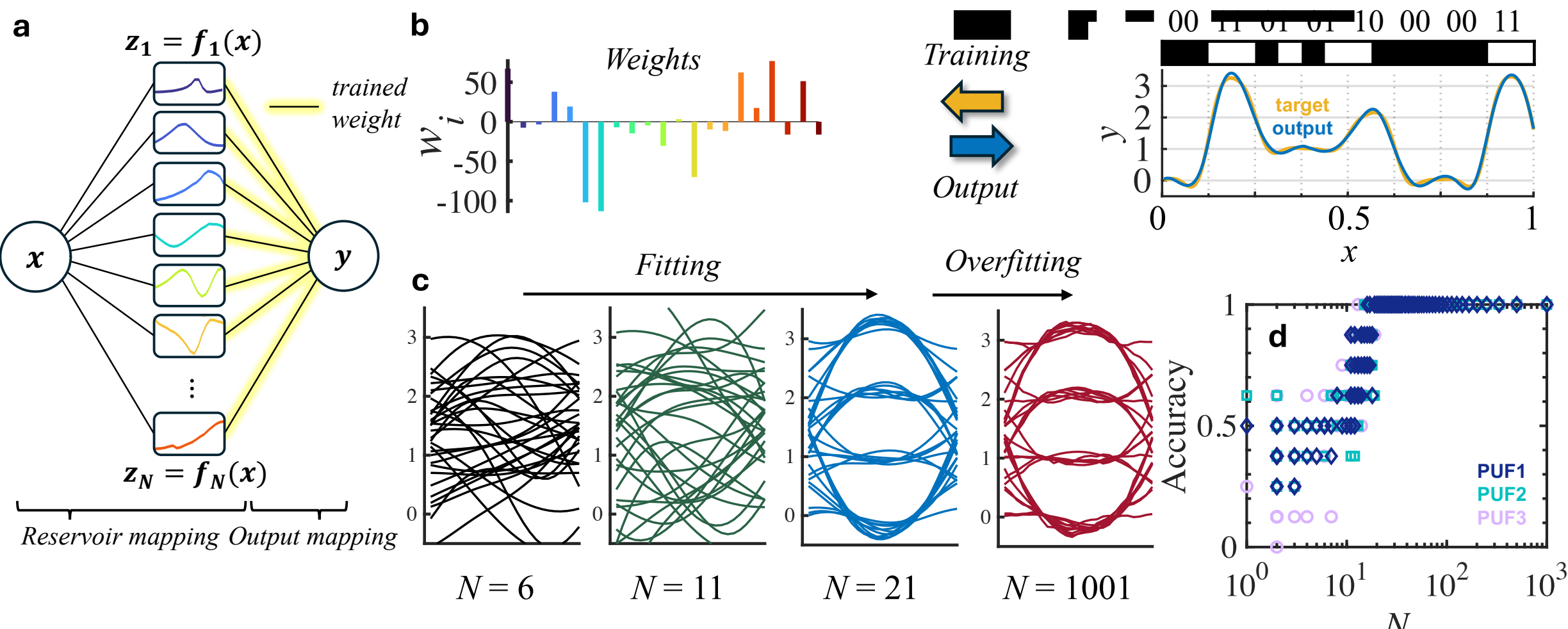

**Figure 6. Reservoir-inspired function approximation with an optoelectronic PUF.** (a) Equivalent neural architecture showing how a target function y(x) is reconstructed from a weighted sum of PUF-derived nonlinear features $z_i(x)$. (b) Example approximation of a 4-level pseudorandom (PAM-4) target function using the trained weights. (c) Eye-diagrams illustrating function approximation performance as the number of nonlinear reservoir features, $N$, is increased. (d) Approximation accuracy as a function of $N$ for three different PUF devices.

## Discussion

Our study demonstrates the dynamic control of physically unclonable integrated photonic circuits based on moiré QCIs subjected to fabrication disorder. These devices challenge conventional paradigms by leveraging low-symmetry and multiple-scattering effects to enhance complex, aperiodic, and reconfigurable spectral responses. In contrast to the analytic behavior of high-symmetry photonic components, the QCI's transient chaos properties support diverse nonlinear feature mappings. These high dimensional mappings exist in the spectral domain, thereby harnessing one of the most advantageous properties of light for encoding information – spectral bandwidth[40].

Recent works have highlighted the ability of linear optics to perform nonlinear information processing[18–21]. Similarly, while optical propagation in our device is strictly linear with respect to the input optical field, the engineered multiple-scattering pathways convert small thermo-optic index shifts into complex interferometric responses. This produces a highly nonlinear mapping between the applied heater voltage and the resulting transmission spectrum. One recent work proposed deep nonlinear encoding based on integrated photonics with deterministically designed arrays of chip-scale resonators[18]. This is akin to intentionally training and controlling the hidden weights **θ** of a complex photonic circuit rather than allowing them to remain random (e.g. Fig. 5a). It remains to be seen whether this is a technologically feasible (efficient and scalable) approach for high dimensional photonic computing, as distributed resonator arrays are inherently fabrication sensitive and would require precise control to restore all weights to their desired values[11]. In principle, sufficiently large resonator networks and MZI networks subjected to fabrication disorder should eventually become PUFs [32,33]. Therefore, it is worthwhile to question whether energy should be expended to tune the internal weights of such a system, as in a conventional deep neural network, or instead expended only on the output layer of a disordered network akin to a reservoir computer.

Along this vein, Xia et al. proposed and demonstrated deep nonlinear encoding based on multiple-scattering in an integrating sphere modulated by a digital micro-mirror device (DMD)[19]. Their architecture leveraged spatial modes at a fixed frequency, whereas our work leverages frequency to achieve high dimensional signal transformation. Although our platform is orders of magnitude more compact, our present architecture is restricted to only two inputs which is substantially less than the $10^2$-$10^7$ afforded by a DMD. Hence, in the future it would be worthwhile to explore randomized neural network processing with integrated photonic PUFs designed to operate with higher dimensional inputs.

On the surface, treating the PUF as a nonlinear processor with an equivalent but unknown neural architecture might appear to imply the PUF is susceptible to modelling attacks. However, this would require an adversary to learn the hidden and inaccessible weights **θ**, that provide the correct mapping $\mathbf{z} = \mathrm{PUF}(\boldsymbol{x}, \boldsymbol{\omega}, \boldsymbol{\theta})$ for all possible values of $\boldsymbol{x}$ and $\boldsymbol{\omega}$**.** Because the PUF operates as a random feature projector, the nonlinear transformation encoded by the internal structure **θ** is fixed and unknown, and only the output weights to $\boldsymbol{\mathcal{F}}(\mathbf{z},\mathbf{w})$ are trainable. While a PUF model could likely be derived that models **z** for subset of $\boldsymbol{x}$ and $\boldsymbol{\omega}$, it would not extrapolate beyond its training data set – and would therefore function more like a look up table attack than a true modeling attack. It remains to be seen if a sufficiently advanced physics informed neural network could accurately model such a PUF. If we conservatively assume such a model is possible to construct, the security of such a PUF would ultimately lie in increasing the dimensionality of $\boldsymbol{x}$ and $\boldsymbol{\omega}$ to construct an exponentially large challenge-response space, thereby making it impossible to collect enough training data to sufficiently map the input-output response in finite time.

## Conclusion

In conclusion, our QCI platform demonstrates how low-symmetry photonic structures can achieve tunable aperiodic spectral responses, facilitating complex input-output characteristics in a compact, chip-scale format. These devices represent practical implementations of localization/delocalization phenomena within disordered, moiré, and/or quasicrystalline photonic structures, offering enhanced and non-redundant mode diversity. Dynamic control of such systems is favorable for realizing transient chaos, expressive nonlinear mappings, physical unclonable functions, and reservoir-inspired processing. This capability opens new pathways for storing, processing, and generating information, offering potential applications ranging from spectroscopy and sensing to hardware security and optical computing.

## Methods

***Device Fabrication:*** Active QCI device fabrication was performed across two fabrication runs of the multi-project wafer (MPW) NanoSOI (220 nm device layer) fabrication process, with oxide cladding deposition (2.2 microns thick) and tri-layer metallization, offered by Applied Nanotools. The tri-layer metallization process introduces 200 nm thick Ti:W alloy (10:90 by weight) thermal heaters with a resistivity of $0.61 \times 10^{-6}$ Ω-cm, with an additional 400 nm of Au on the routing layer, and a 300 nm thick $SiO_2$ passivation layer. Each QCI resistive heater has a resistance of 580 Ω.

***Numerical Simulations:*** The numerical simulations reported in Fig. 2 were performed using the transfer matrix method described in the supplementary information of Ref. [15]. Briefly, the moiré quasicrystal is modeled in 1D using a waveguide effective index profile $n_{eff}(s)$ defined along the optical path length $s$. A complex effective index profile is used to account for the measured ~2.4 dB/cm propagation loss of a straight 500 nm wide waveguide. To emulate nanoscale fabrication variations, small random perturbations (Gaussian-distributed) are added independently to the real and imaginary parts of $n_{eff}(s)$ representing local width fluctuations and surface-roughness-induced scattering. The electric field intensity vs. length is extracted according to the approach utilized in Refs. [41,42].

***Opto-electronic Device Measurement:*** A preliminary sample was studied in the Nanophotonics Laboratory at Clemson University using a tunable wavelength laser source (Santec TSL-510), polarization controller, fiber circulator, photodetector, electrical probes, wirebonding, and a dual channel current source (Keithley 2602B). The final sample was characterized at The University of British Columbia using a custom-built electro-optic test-station. The test-station supported computer-automated grating coupled spectral measurements with a tunable laser (Agilent 81600B) and optical power meter (Agilent 81635A) to capture device spectra over the range 1480–1580 nm in 10 pm steps. Electrical control was driven by the SiEPIC Lab control suite and a dual channel source-measure unit (Keithley 2604B) in which a wedge probe, controller on piezoelectric actuator, is used to make contact to the electrical pads on the photonic integrated circuit.

**Supporting Information**

The supporting information includes supporting videos and figures. Supporting Videos 1-4 show animations of 2D state space trajectories for analytical ring resonator and rPUFs 1-3. Supporting Figures SFigs. 1-4 show additional spectral characterization and analysis of optoelectronic QCIs. SFigs. 5 & 6 show additional function approximation examples.

**Acknowledgement**

This research was supported in part by the Air Force Office of Scientific Research (AFOSR) under grant no. FA9550-19-1-0057 and the National Science Foundation (NSF) under grant nos. 2235443 & 2413234. The authors thank Derrick Joyce for aiding with preliminary experiments. L.C. acknowledges support from The Canada Foundation for Innovation (CFI) and Natural Sciences and Engineering Research Council of Canada (NSERC).

**Author Contributions**

Y.L. and J.D.R conceived the project. F.B.T. lead the design, initial experiments, and analysis. M.H., J.B., M.M. and L.C. contributed to the development of the electro-optic experimental setup and J.B. performed the final measurements. F.B.T. and J.D.R analyzed and interpreted the results. Y.L., T.K., and L.C. assisted in the interpretation and analysis. F.B.T. and J.D.R. prepared the figures. J.D.R. supervised the project and drafted the manuscript with feedback from all authors.

**Ethics declarations**

The authors declare no competing interests.

**Data Availability**

The data that support the findings of this study are available from the corresponding author upon reasonable request.